# Analog Programing of Conducting-Polymer Dendritic Interconnections and Control of their Morphology

Kamila Janzakova[a], Ankush Kumar[a], Mahdi Ghazal[a], Anna Susloparova[a], Yannick Coffinier[a], Fabien Alibart[a,b] & Sébastien Pecqueur[a]*

[a]Univ. Lille, CNRS, Centrale Lille, Univ. Polytechnique Hauts-de-France, UMR 8520 - IEMN, F-59000 Lille, France.

[b]Laboratoire Nanotechnologies &Nanosystèmes (LN2), CNRS, Université de Sherbrooke, J1X0A5, Sherbrooke, Canada.

Email: sebastien.pecqueur[at]*iemn*.fr

## Abstract

**Although materials and processes are different from biological cells', brain mimicries led to tremendous achievements in massively parallel information processing via neuromorphic engineering. Inexistent in electronics, we describe how to emulate dendritic morphogenesis by electropolymerization in water, aiming *in operando* material modification for hardware learning. The systematic study of applied voltage-pulse parameters details on tuning independently morphological aspects of micrometric dendrites': as fractal number, branching degree, asymmetry, density or length. Time-lapse image processing of their growth shows the spatial features to be dynamically-dependent and expand distinctively before and after forming a conductive bridging of two electrochemically grown dendrites. Circuit-element analysis and electrochemical impedance spectroscopy confirms their morphological control to occur in temporal windows where the growth kinetics can be finely perturbed by the input signal frequency and duty cycle. By the emulation of one of the most preponderant mechanisms responsible for brain's long-term memory, its implementation in the vicinity of sensing arrays, neural probes or biochips shall greatly optimize computational costs and recognition performances required to classify high-dimensional patterns from complex aqueous environments.**

# Introduction

Brain-inspired computing has reached several important milestones in the last decades that are positioning it as a realistic alternative to the limitation of Turing's paradigm in the big data era.[**Hassabis2017**] As hardware, information processing technologies are intrinsically bound to materials that are chosen to carry quanta of information, and fabrication processes to shape them into dense structures of computing device elements.[**Kusum2013**, **VdBurgt2017**, **Sangwan2020, Gerasimov2019**, **Fuller2019**, **Tang2019**, **Pecqueur2018**] In software, Machine-Learning based Artificial Neural Networks (ANN) are algorithms that have demonstrated above human-level performances for a large variety of high-dimensional data processing tasks such as image,[**Fadlullah2017**] or speech recognition.[**Nassif2019**, **Hinton2012**] These achievements are currently enabled by the level of performances provided by modern computers that allow processing ANN models over realistic durations for their practical application. However, running such algorithms on current computers is highly energy-demanding comparatively to the brain, up to major concerns for applications as embedded computing for robotics or sensing for brain-machine interfacing. If important improvements at the algorithmic level certainly have to be done to reach brain-like performances in terms of power consumption, the hardware that supports such unconventional computing remains an open field for innovation in that sense.[**Davies2018**] One particular aspect that conventional electronics do not exploit is *in operando* device bottom-up engineering. On the contrary, such equivalent processes naturally occur in the brain by the form of diverse biogeneses that create physical connections, defining optimal computing topologies or network self-healing. In particular, brain's dendrite morphogenesis originates on each neuron upon the whole life of a learning system, and controls each neuron's growth as a full functionality for the brain's long-term memory.[**Prigge2018**] In conventional hardware, devices are manufactured in a top-down fashion prior their use, and all possible electrically-erasable connections and network parameters need to be defined *a priori*. While the biological approach maximizes resources' utilization, the electrical one requires oversizing network dimensions. Exploring the possibility of its embedding in electronic devices would provide disruptive solutions and perspectives for hardware ANN implementation. To this end, this study focuses on electropolymerization to form electrical connections between nodes, at a surprising mimicry level with biological dendrites, with enough versatility to tune their morphology by the input voltage-pulse dynamics that activates the hard wiring. Neural dendritic patterns in the brain can be classified into thirteen families of three groups of dendritic fields,[**Abdel-Maguid1984**] and each families of dendritic arbors have specific characteristics of branching number, number of segmentation per dendrites and the length and volume of the different branches and segments.[**Bicanic2017**] As dendrites' morphology contributes in the electrophysiological features that yield their signaling,[**Grudt2002**] the control of this versatility is essential to tune the signal propagation associated to the temporal characteristics of their memory property. Our study investigates on the input voltage pulse parameterizing to grow electrical dendrites and on how to changes their shape.

## Results & Discussion

Earlier results on AC bipolar electropolymerization of 3,4-ethylenedioxythiophene (EDOT) have validated the oxidative formation of electropolymerized PEDOT (ePEDOT) dendritic structures on gold wires at the millimeter scale, supported by the reductive counter-reaction of benzoquinone (BQ) into hydroquinone (HQ).[**Watanabe2018**, **Ohira2017**, **Koisumi2018**, **Inagi2019**, **Koisumi2016**, **Eickenscheidt2019**] Some of the advantages of such mechanism leading to the conductive bridging of polarizable nodes in a common electrolyte are local-activation (for individual programing), dependency on electric field's direction (for layer-to-layer interconnection parallelism), dependency on electric field's magnitude (for neural convolution) and voltage non-linearity of the electropolymerizations activation (for neural rectification). The study focuses on the dendritic growth of EDOT with BQ upon the polarization of two free-standing gold wires in an electrolyte (Fig.1a). To keep this study as systematic as possible and not observing morphological control of the dendrite due to the environment, cleaned gold wires have been lifted up to the same height above the same electrolyte-supporting substrate. The wire ends have been kept still and distant by L = 240 µm. We confirm the reaction to be field-dependent as we observed that wires' vicinity can either enable or disable of the dendrite formation without changing the applied voltage. In this study, water was used as the only solvent to support sodium polystyrene sulfonate (NaPSS) as electrolyte.[**Schweiss2005**] Although wires were polarized at larger voltage biases than water's electrochemical window, we confirm that water electrolysis is not preponderant during our dynamical operations: First, it was observed that lack of $BQ_{(aq)}$ results in the dendrite formation disabling. Second, the growths of ePEDOT dendrites were not accompanied by any apparent bubbling on the wires, within the applied voltage boundaries of our experiments (except only in the last experimental part of our study on the duty cycle). $EDOT_{(aq)}$ and $BQ_{(aq)}$ were always initially introduced equimolar at 10 mM, limited by the moderate solubility of EDOT in water.[**Schweiss2005**] In this study, growths were performed by applying a pulse voltage waveform at the signal wire and grounding the other wire, such that the applied voltage bias is defined by (Fig.1c):

$$V_{sig}(t) - V_{gnd} = \begin{cases} V^{+} for\ t \in [0; dc * T[\ mod(T) \\ V^{-} for\ t \in [dc * T; T[\ mod(T) \end{cases}$$

With $V^{+}=V_{off}+V_{pp}$, the anodic polarization applied at the signal electrode versus ground ($V_{pp}$ defined as peak voltage amplitude and $V_{off}$ the voltage offset) applied for each period T=1/f between 0 and dc*T (dc being the duty cycle) and $V^{-}=V_{off}-V_{p}$, the applied cathodic polarization at the signal electrode versus ground, applied for each period T between dc*T and T. As the applied voltages are no electrolyte-potentiostatic, the electrochemical potential drops between both interfaces from $V'_{sig}$ at the signal wire dendrite and $V'_{gnd}$ at the ground wire one (points respectively depicted in orange and purple in Figure 1b). $NaPSS_{(aq)}$ was chosen as the electrolyte, for the perspective to form highly doped ePEDOT dendrite of conductivity up to 1 kS/cm [**Rivnay2016**]. By considering the redox couples on both wire/electrolytes interfaces separated by the ion-conducting medium, such dynamical system obeys to an equivalent circuit, idealized in Fig.1b (assuming no transfer-limiting preponderant phenomena, leading to constant-phase electrochemical elements)[**Orazem2017**]. The model considers the electrolyte conductance $G_{Na+PSS-}$ with $Na^{+}_{(aq)}$ and $PSS^{-}_{(aq)}$ drifting under voltage polarization. Both kind of ions accumulate under steady-state at the electrolyte/dendrite interface of the opposite polarity to form a dense double layer of $PSS^{-}_{(aq)}$ on the cathode (capacitance $C_{PSS-}$) and of $Na^{+}_{(aq)}$ ions on the anode ($C_{Na+}$). The respective double-layers permeation to neutral solutes rely on their packing, modulating the Faradaic processes of EDOT oxidation (resistance $R_{ox}$) and BQ reduction ($R_{red}$) during the transient. Depending on the analytical expression of the applied dynamic voltage waveform, processes are periodically reversed between both signal and grounded electrodes. Considering ions electrostatic relaxations in the equivalent circuit simplified by fixed ohmic resistors R and dielectric capacitors C, the dynamics of the surface potential on both wires can be predicted as depicted in Figure 1d (considering the electrolyte to be comparatively as resistive as the charge-transfers, and the dynamics of the wires' RC charge and discharge to be comparable to the period of the applied signal). We studied iteratively the influence of the four voltage/temporal waveform parameters, ruling the dynamics of the wires surface potential, namely the voltage amplitude, the voltage offset, the frequency and the duty cycle of the waveform.

**Morphological influence by the peak voltage amplitude:** The electrochemical evaluation of the 2-wire system under cyclic voltammetry (CV) shows that a minimum voltage as low as $\Delta V_{min} = 0.9$ V is required to electropolymerize $EDOT_{(aq)}$ with $BQ_{(aq)}$ under quasi steady conditions (see Fig.2a – 50 mV/s). Being twice lower than in acetonitrile,[**Koizumi2016**] these potential shifts can be attributed to water as protic solvent, which influences both the kinetics (redox process involving proton transfer) and the thermodynamics by the different species solubility due to hydrogen bonding.[**Quan2007**] $PSS^{-}_{(aq)}$ may also influence the stability of the products both as a surfactant and as a dopant that promotes the ionization of the ePEDOT semiconductor in water. CV shows that the $NaPSS_{(aq)}$ electrolyte is not electrochemically active in this investigated potential, which supports that $BQ_{(aq)}$ reduction is the counter reaction for dendritic ePEDOT electropolymerization, thus does not require the participation of $PSS^-$ as redox active component.

Then, by electropolymerizing with peak amplitude voltages $V_p \geq 3.5$ V, we observe nucleation on both wires and growth of material at the minute scale (Fig.2b at $V_p = 3.5$ V, $|V^+| = |V^-| = 3.9\ \Delta V_{min}$), without apparent bubbling at any of both wire electrodes. This indicates that during the transient process, charge-transfer reaction at both electrodes are not thermodynamically driven (f = 80 Hz, dc = 50% in Figure 2) as such large voltage biases would also promote water electrolysis under steady conditions. The specific morphology of the growing polymer do not show particular regularity along the wire-to-wire axis, but protuberances with spatial periodicity that do not seem to be directly correlated with the temporal periodicity of the applied waveform (Fig.2b). The fiber-like ePEDOT objects shows to be mechanically fragile as it could hardly be pool out of the aqueous electrolyte without damage. Scanning electron microscopy (SEM - Fig.2c-e) shows their brittleness with sharp-edged cut on the pooled-out object, that suggests non-plastic properties of the polymer at these dimension scale. This evidences strong polymer chain interactions, led by either strong π-π stacking between the lipophilic PEDOT chains in the water solvent, and/or strong ionic bonds between ePEDOT and $PSS^-$ ionomer chains compounded inside the dendritic structures. Steady electrical tests show conductivity at the order of 0.1-1 S/cm (Fig.S1 as a supplementary material), which is only 3-fold lower than the one of PEDOT:PSS optimized dispersions,[**Rivnay2016**] and in the conductivity range of other polythiophenes with high-doping level.[**Ferchichi2020**] This suggests that $PSS^-$ is included into the ePEDOT matrix and a partial doping of the semiconductor material. SEM pictures also show that these objects gather multi-scale morphological richness by their apparent branchy structure at the 10-100 µm range, and also sub-µm rough texturing (Fig.2d-e). To study on the higher-scale morphological control of these structures, a first voltage-parametric investigation was carried out by the variation of $V_p$ from 3.5 V to 6.5 V at f = 80 Hz, dc = 50% and $V_{off} = 0$ V. We observed first that the dynamics of the growths was highly depending on $V_p$ (Fig.2f-h): From the time-lapse frame processing of the microscope video recorded during the growth at different $V_p$, the projected area (2D image of the 3D object) was increasing with time as the ePEDOT grows on both wires, with a strong dependency on $V_p$ (Fig.2f). As the waveforms are statistically symmetrical around 0 V with equal electric fields on both a half-period (dc = 50%, $V_{off}$ = 0 V) the material grows evenly (in size and morphology) on both wires in each $V_p$ case (Fig.2h). The growth rate is very slow at voltages from $V_p$ = 3.5V, and saturates at $V_p$ = 5 V with a rate of 150 µm²/s. This indicates that at least one of the charge transfer rates for either oxidation or reduction reaches an upper limit by the activation with the voltage, yielding growths that are sensitively similar for $V_p \geq$ 5V under these conditions. From the discrete 1D-spatial decomposition of the 3D dendrite into elementary cylinder (details in Figure S2 as supplementary information), this upper limit for the projected rate is estimated to a volume rate of 3.68 $10^3$ µm³/s, corresponding to a deposition rate of 5.1 ng/s (assuming 1.4 g/cm³ as PEDOT-based polymer density).[**Chanthaanont2013**, **DeLongchamp2005**, **Wei2016**, **Wei2015**] Assuming an apex-growth over a specific area comprised between one and ten times the gold wire orthogonal cross-section, the surface rate is estimated to be at the order of the 0.1-1 mg/cm²/s (equivalent to the oxidation of 4 to 44 EDOT molecules per nm² per nanosecond), or corresponds to 750-7500 nm/s in the equivalent case for the coating in a homogeneous thin-film. While this rate is at minimum 15 times higher than what is experimentally observed in potentiostatic electrodeposition of ePEDOT thin-films (we usually obtain only up to 50 nm/s for $EDOT_{(aq)}$ in $NaPSS_{(aq)}$ at 1.5 $V_{DC}$ bias between two Au electrodes), this suggests a higher turbulence in the electropolymerization mechanism under such waveforms (caused also by the periodic modulation of both anions and cations Helmholtz layer on each wire), that probably lead to dendritic arborescence rather than films under such conditions. As the growth rates differ with $V_p$, they also converge at different completion times from 178 s at $V_p$ = 5 V to 423 s at $V_p$ = 3.5 V (Fig.2f,h), systematically at the mid-distance of the inter-wire gap, to form a conductive bridge between both gold wires (Fig.S1). The morphology of these dendrites are also $V_p$ dependent, as low-$V_p$ growths appear as lower fractality object with branching degree close to one, while higher-$V_p$ objects appear to be more dendritic with an apparent branching degree up to five (defined as the average number of parallel branches on all longitudinal positions between the gold wires). The branching factor and fractal dimension increases from 3.5 to 5 V and decreases afterwards. The decreasing part may be due to higher density of dendrites in a limited area, reducing the surface fractal nature and number of branches. During all growths with $V_p$ variation, new ePEDOT nuclei was systematically favored to be seeded on the apex of freshly grown branches rather emerging than on the periphery of the gold wires. This is characteristic of two properties: First, the field-effect activation of the electrochemical process, as in each case the dendrites appear to grow preferentially on the longitudinal wire-to-wire axis than out of it). Second, it shows a lower charge transfer rate at the gold/electrolyte interface than the ePEDOT/electrolyte one (at least for EDOT oxidation), which testifies on both the low resistance of the gold/ePEDOT contact and good conductivity of the ePEDOT material. In this setup, the field-effect sensitivity (that was experimentally evidenced by the sensitivity of the growth to the gold wire spacing, but is not tediously investigated in this study) is justified by the influence of the anode-to-cathode distance that controls the ionic transport resistance 1/G, that in turn control the voltage drop at both anode/electrolyte and cathode/electrolyte interfaces.

**- Morphological influence of the voltage offset:** By introducing an offset voltage component ($V_{off}$) in the pulse waveforms, a strong morphological asymmetry was promoted in the dendritic growth on both wire (Fig.3). The $V_{off}$ parameter can influence the system by two distinctive ways: First, $V_{off}$ induces an asymmetry in the applied oxidation and reduction voltages simultaneously experienced on both wires, which in turn influences the electropolymerization rate individually on the wires. Second, $V_{off}$ generates also different electric-fields applied during both half-periods, for which the field lines shall govern the growths orientation. From their cumulative effects, we distinguish both the growth rate and the growth anisotropy can be influenced by $V_{off}$ and result in the different dendritic morphologies. Experimentally, we observed that the sign of the offset control morphological asymmetries in both dendritic growths: the larger arborescence is systematically observed on the signal wire with positive offsets, whereas it is always the ground for negative offsets, with a monotonic trend in the gradual asymmetry control with $|V_{off}|$ (Fig.3a,c). The time scale for the dendrites formation decreases with the increase of $|V_{off}|$ (Fig.3b,c). Despite the offset voltage induces electric-fields with different magnitudes on the dendrites forming at both wires, both anodic growths always meet specifically at their mid-gap between the gold wires. This shows that although the offset promotes the dendritic growth on one electrode over the other, the equatorial component for both growths (parallel with the electric field) remains the same, with an equal growth velocity that is not correlated to the dendrite morphology. Morphologies for both dendrites are governed by distinctive growing modes: the thinnest dendrite's appears to be the most linear and longitudinal with the equatorial direction, while the densest one shows always radial growth (Fig.3b). This evidences that the thinnest dendrite (experiencing lower applied oxidation voltage) is more dominated by the electric field than the thickest one (experiencing higher applied oxidation voltage). Therefore, dendritic growths are independently governed by both the voltage amplitude at the electrode and their induced electric field. As depicted in Fig.1c, wires behave both as anode and cathode in the whole experiment (if $V_p > V_{off}$, as in this experiment) and dendritic growth is indeed expected to occur on both wires (as in all cases here $V_p$-$V_{off} > \Delta V_{min}$). However, as the wire experiencing the highest oxidation stress shall expose also to the highest field polarization, this show that field orientation controls the growth only for low growth rates, as the thin dendrites appear straighter along the field orientation than thicker ones. The experimentally observed change of dominance in growth orientation is evidencing voltage non-linearity in the charge-transfer yielding anodic processes of electropolymerization on the wire. Non-Ohmicity for charge-transfer resistances under quasi-steady conditions was evidenced by voltage-ramped impedance spectroscopy (Fig.3f). As the direct-current voltage bias ($V_{DC}$) is gradually increased over time in the experiment, we observed a sudden decrease in the charge-transfer resistance $R_{CT} = R_{ox}+R_{red}$ when approaching $V_{DC} = \Delta V_{min}$, increasing with $V_{DC}$ after reaching a minimum in the MΩ range (Fig.3f). Confirming the voltage activation of the growth, it also invalidates the linearity of the applied electric-field $E = (V'_{sig}-V'_{gnd})/d$ (d<L as shortest distance between two dendrites), with $V'_{sig}$ and $V'_{gnd}$ being $R_{CT}(V_{DC})$ dependent, and therefore dependent on $V_{off}$. Furthermore, we observed the growth continued further after contacting each other, by the appearance of radial dendrites on the thickest dendrite wire for higher $V_{off}$ values (easily evidenced in Fig.4b for time lapses at $V_{off}$ = -0.4 V, +0.2 V and +0.8 V recorded until $t_{end}$ longer than their respective completion time). Such feature indicates on the good conduction for these asymmetric objects after merging: As none of the transmission lines where loaded with serial resistances on any sides (other than the 50 Ω protection of the generator and the < 1 Ω of our coaxial setup), the electrical conduction is limited by the resistance of the fused dendrites, for which the higher value causes the voltage to drop all along both dendrites. Leaving the highest point of the applied potential to be strictly on the gold surface of the anodic wire, electropolymerization continues further after merging by the creation/expansion of new branches on gold.

**- Morphological influence of the frequency:** The influence of the frequency using square waveforms on the growth of ePEDOT wires in a water/acetonitrile electrolyte has recently been shown,[**Akai-Kasaya2020**] and evidences an empirical scale law existing between the apparent diameter of the wire-like ePEDOT structures and frequency of the applied voltage waveform. Here, we focused on a lower frequency range (between 10 Hz to 1 kHz) to monitor the morphological change transition from dendrite-like to wire-like that occurs by the increase of the frequency in water (Fig.4). As the electrochemical system starts to show bubbling for frequency only below 10 Hz (at least experimentally evidenced at 1 Hz), this property orientates the earlier conclusion that the redox activity of the system is kinetically limited, as the thermodynamically favored water electrolysis is enabled if the applied voltage remains still for a sufficiently long time (from the 100 ms range).

The frequency variation in the voltage pattern shows a relationship between the projected area and the frequency that appears inversely proportional (as shows the -1 slope in double log scale in Fig.4b), suggesting here that projected area is linear with the waveform's period. Since the projected area is directly proportional to the dendrite's branches diameter by the elementary cylinders approximation (at least for the highest frequency cases where dendrites appear as linear cables), the quadratic low identified for lithographically patterned Au-electrodes in acetonitrile/water at higher frequency appears linear for our experimental conditions for dendritic growth. At higher frequency, dendrites' growth seems also to take shorter to merge, by a linear relationship between completion time and frequency displaying a -1 slope in double log scale (Fig.4b) at frequency below 200 Hz. In case of lower branching dendrites above 200 Hz, we observed dendrites to grow slower, by a -1/2 slope on the double log plot in Fig.4b. However, we did observe a deviation from this trend at frequencies higher than 200 Hz, where dendrites are substantially much thinner: dendrites merge much slower up to 950 Hz. Nuclei with micrometer-scale branches at frequency above 1 kHz were observed but have not been attempted to merge at this inter-wire distance, due to a too long completion time, not compatible with preserving the electrolytic drop from substantial evaporation. From the impedance spectroscopy of three different dendritic doublet morphologies (thin and vicinal, thick and vicinal and thick and distant – details in Figure S4 as supplementary information), all three pairs of dendrites have a comparable spectral behavior with a cut-off at around $f_c$ = 20 Hz. Compared to naked gold wire, all these dendritic systems take longer to charge their double layers (observed cut-off between 1 and 10 kHz for naked gold in Fig.3e). With dendrites' specific areas being much larger than the wire, the increase of the double-layer capacitance at the electrochemical interfaces is responsible for the lower frequency shift of the cut-off. Providing that this 20 Hz cut-off is similar for all three pairs of dendrites, we observe that the first 25 ms of the wire polarization appears to be a specific time window in the charge transient of dendrites of these sizes. Earlier than $t=1/2f_c$, the ionic space-charge around the electrodes is not fully-formed as compact Helmholtz layers to screen the voltage polarization at the electrodes vicinity in the electrolyte ($V'_{sig}$ near the signal and $V'_{gnd}$ near the ground, as depicted in Fig.1b). Assuming ideal resistor-capacitor charge/discharge for the $PSS^-_{(aq)}$ Helmholtz layer at the anode, a significant exponential decay of $V'_{sig}(t)$ in the millisecond timescale, which therefore conditions the oxidation rate of $EDOT_{(aq)}$ in the sub-kHz range (Fig.4c). As the correlation between spectral dependency growth morphology is not straightforward, the mechanistic interpretation of the dendrites' growth morphology with voltage frequency remains only qualitative. However, in light of the spectral and image analysis, we can assess that the process is not limited by an anodic monomer gradient diffusion, as the most massive dendrites are obtained at the lowest frequency for a same polarization duration. As in this case the temporal density of anodic polarization equals 0.5 (dc = 50%) for all frequency conditions, the densest dendrites would have been expected for the highest frequency in case of diffusion limitation. Also providing the ionomeric structure of the $PSS^{-}_{(aq)}$ polyanions, one can assume that its time-dependent accumulation on the ePEDOT surface screens the dendrites both electrostatically but also sterically. Such physical hindrances for the $EDOT_{(aq)}$ oxidation can potentially justify the observed gradual transition of morphology from isotropic and diffused growth to oriented and linear ones (Fig.4d).

**- Engulfing asymmetry of the duty cycle:** The duty cycle parameter is the parameter that differentiates the square waveforms among the other pulse ones, as specific cases in which dc = 50%. By varying dc value by ±10% around the one of a square waveform, a comparable trend was observed as previously with the variation of $V_{off}$, as the asymmetry favors the dendritic electropolymerization on one wire over the other (Fig.5a): At dc = 60%, the signal's dendrite is denser than the ground's and vice versa at dc = 40%, as the growth is promoted on the wire that experiences the longest anodic polarization duration in the waveform expression. Such close relationship between dc and $V_{off}$ can also be observed in the Fourier decomposition of the pulse waveform (eq.1 and eq.2),[**Smith1997**]

$$V_{sig}(t) - V_{gnd} = V_{off} + (2dc - 1)V_p + \sum_{n=1}^{\infty} \frac{4V_p}{n\pi} \sin(\pi\, n\, dc) \cdot \cos(2\pi\, n\, f \cdot t) \qquad \textbf{(eq.1)}$$

$$V_{sig}(t) - V_{gnd} = dc\, V^+ + (1 - dc)V^- + \sum_{n=1}^{\infty} \frac{2(V^+ - V^-)}{n\pi} \sin(\pi\, n\, dc) \cdot \cos(2\pi\, n\, f \cdot t) \qquad \textbf{(eq.2)}$$

Where introducing an asymmetry of the temporal polarization densities by dc allows inserting a time-invariant term $V_{DC}$ in the waveform without breaking the applied-voltage asymmetry of the $V^+$ and $V^-$ around zero, such that (eq.3):

$$V_{DC} = V_{off} + (2dc - 1)\, V_p = dc\, V^+ + (1 - dc)\, V^- \qquad \textbf{(eq.3)}$$

From the Fourier transform expression of the pulse waveform, the $V_{DC}$ component introduced by the temporal unbalance of the polarization equals a sum of $V_{off}$ and a fraction of $V_p$, such that a dc = 60% generates $V_{DC}$ = 1 V, 2 V at 70%, -1 V at 40% and -2 V at 30% (for $V_p$ = 5 V, $V_{off}$ = 0 V). The dependency of the projected area ratio signal/ground $\rho(dc, V_{off})$ with the duty cycle illustrates this relationship in our experiments: at completion, time ρ(40%,0V) = 0.18 and ρ(60%,0V) = 3.11, while previously we observed ρ(50%,-0.8V) = 0.24 and ρ(50%,+0.8V) = 3.1 (Fig.3c), suggesting that the dc component in the waveforms rules the dendrites' asymmetry in both $V_{off}$ and dc variation series. This evidences that the electropolymerization experimentally occurs specifically when the double layer charges and not in the electrolytic conduction regime (that is independent from $V_{DC}$). This further confirms the conclusion of the frequency variation, showing that a frequency increase results in a lower projected area). In terms of circuit element analysis, the control of the $V_{DC}$ bias by the dc originates from the temporal asymmetry of the charge/discharge processes at both electrode, which results in the partial charging of one specific type of ionic charge carrier on both electrodes favored by the longest time polarization. Specific to the dc (previously $V_p$, $V_{off}$ and f did not promote such effect), its asymmetry creates a charge build-up at the double layer capacitors of both dendrites that induces a dc-dependent serial potential pinning. From the electric simulation of the first-level approximated equivalent circuit previously considered (Fig.3f inset), one can observe that such a charge build-up induces a transient dc-dependent over-potential at both electrode interfaces that can be higher in absolute value than the applied voltage (see Fig.5b). At the beginning of the experiment (t=0 in Fig.5b), both double layers are unformed around the wires, so the initial state of their corresponding capacitor are fully discharged and induce exponential decays with $V_{sig}$ and $V_{gnd}$, as starting values of their surface potentials. At the second change of polarization ($t=T^{+}$=dc*12.5ms in Fig.5b), one can observe the electrostatic charge/discharge of the double layer capacitors initiate from different values than $V_{sig}$ and $V_{gnd}$ because of the partial charge by past polarizations. So in Figures 5c, we observe that $|V'_{sig}|>\max(|V^{+}|;|V^{-}|)$ and $V_{gnd}\neq 0$ at the polarization inversions in the applied voltage wave, and therefore the generated transient electric-fields across the electrolyte are much higher in magnitude than the external $V_{sig}/L$ applied on the gold wires (insets in Fig.5c graphs). While this phenomenon is commonly induced in each of the previously studied cases, the asymmetry on the duty cycle allows stretching the time windows for one preferential ionic charge/discharge over the other at each electrode, which in turns modifies the initial potential pinning of the initial state of the charge/discharge at each potential inversion. One can notice in Fig.5b that although the duty cycle does not modify the decay rate by decreasing dc from 50% to 30%, it changes the initial values from +5.9V/-5.9V for $V'_{sig}$ and +0.9V/-0.9V for $V'_{gnd}$ at dc=50% to +6.7V/-4.9V for $V'_{sig}$ and -1.7/-0.1V for $V'_{gnd}$ at dc=30% (Fig.5b,c). As in eq.3, this effect is linear with the change in duty cycle (Fig.5c), and can even reach a state where the surface potential at the ground electrode keep the same polarity vs. gnd over time (see dc=30% in Fig.5c, where max($V'_{gnd}$) crosses the V=0 line): Although one gold wire is grounded at $V_{gnd}$=0V, its surface potential $V'_{gnd}$ at its dendrite/electrolyte interface under strong $T^+/T^-$ asymmetry creates a permanent formation of exclusively one kind of ion. In Figure 5c, the simulation shows also that the voltage drop $V'_{sig}$-$V'_{gnd}$ across the electrolyte at the polarization change is strongly asymmetric by varying dc, with the initial values $(\max[V'_{sig}]-\min[V'_{gnd}])_{Vsig=V+}/(\min[V'_{sig}]-\max[V'_{gnd}])_{Vsig=V-}$=+6.9V/-6.9V at dc=50% and +8.3V/-4.8V at dc=30%. This implies that under an asymmetric $T^+/T^-$ regime with the dc, the thinnest dendrite grows under a larger electric field than the thickest one (insets of Fig.5c), which agrees with the experimental observation that the thinnest dendrites seem more unidirectional along the inter-wire axis than the bulkiest radial dendrites, grown during a same experiment (Fig.5a). Finally, it was also experimentally observed that the dendrites grown under such an asymmetric $T^+/T^-$ regime have a particularly peculiar behavior after completion time, as they continue to propagate along the wires and densify the connection in polymer material (Fig.5d). In the previous experiments with dc=50% for the study of the influence of $V_p$, $V_{off}$ and f, all dendrites converged to the mid gap with difficulty to merge: a behavior that is particularly obvious with the $V_{off}$ variation in Fig.3b and where we observed that the growth after completion time creates new branches on the gold edge. Instead, we observe here that biasing the duty cycle allows to formed connections that appear more robust, with the two wires welded together by the thickest ePEDOT dendrite engulfing the thinnest one after completion time. While the reason has not been clearly identified yet, this particular property to modify the conductance of the dendritic connection shall greatly been advantageous to exploit in the future as a versatile parameter in a larger structure as a neuromorphic network, to program the conductive interconnections fed as a spiking neural network.

# Conclusion

Dendritic electropolymerization of EDOT monomer in water has been evidenced here to be an electrochemical process that mimics neural dendrite arbor morphogenesis with a large versatility to tune electrical interconnection properties. Voltage offset, amplitude, frequency and duty cycles of applied voltage spikes promote specifically the growth of a conducting polymer bridges. Dendrites' branching number, orientation, asymmetry, engulfing, surface and length of the dendritic segments are ruled by the spike parametrizing, and the different electrochemical processes undergoing at different time domains on the different interfaces. Morphologies of such conducting bridges yield the impedance property by specific frequency-dependent resistance and capacitance that engrave the experienced voltage-pattern in a unique-topology dendritic-interconnect. Narrowed to voltage parametrizing that are compatible with 5V logic and frequencies far below any microcontroller clocks currently used in digital electronics, this study shows the relevance of such an electrochemical process for its integration in the largest electronic application field. Applied as an analog-front-end to encodes amperometric/conductimetric sensor signals into hardware morphogenesis that cannot be reverse-engineered, dendritic electropolymerization as neuro-inspired process shall greatly decrease the vulnerability of computing hardware systems that requires both higher level of security and computing power to process personal data in new biometric sensing technologies.[**Gao2020**]

# Methods

**Materials and instrumentation.** Dendrites were grown by bipolar alternating current electropolymerization technique in an aqueous electrolyte containing 1 mM of poly(sodium-4-styrene sulfonate) (NaPSS), 10 mM of 3,4-ethylenedioxythiophene (EDOT) and 10 mM of 1,4-benzoquinone (BQ). All chemicals were purchased from Sigma Aldrich and used without any prior purification. Two 25 µm-diameter gold wires (purchased from GoodFellow, Cambridge, UK) were employed as working and grounded electrodes, immersed into a 20 µl drop of electrolyte placed onto a Parylene C covered glass substrate. Both gold wires were systematically lifted-up on a controlled height from the substrate and placed at a distance of 190-200 µm from each other. Square-wave signals were generated from a 50MS/s Dual-Channel Arbitrary Waveform Generator (Tabor Electronics), with consistent study of electrical parameters' impact: peak amplitude voltage ($V_p$), frequency (f), duty cycle (dc) and offset voltage ($V_{off}$). Each dendrite growth was carried out with unused gold wires and daily prepared solutions.

**Image processing.** Growths were recorded with a VGA CCD color Camera (HITACHI Kokusai Electric Inc). Video pre-processing involves greyscale-256 at 1 fps data conversion (with the VirtualDub open-access software), prior single-frame image processing (with the ImageJ open-access software). Each frame during the dendritic growths was linearly transformed with the StackReg and TurboReg ImageJ addons,[**Thévenaz1998**] to compensate the time-dependent optical aberration caused by the electrolyte drops changing its curvature overtime due to partial evaporation. Furthermore, frames were cropped around the dendrites and binarized black/white with an appropriate user-defined contrast setting for the different videos generated for each growth. Time-lapses were color encoded with the corresponding ImageJ built-in tool to encode the dynamics of the growth on a scale from dark purple (for the beginning of the growth), to light yellow (for the end of the growth).

**Feature extraction.** The image parameters were calculated in Python, involving binarization, pixel counting, fractal dimension determination, branch calculation and volume determination for the different electrical conditions. Dendrite density was obtained based on the pixel counting from the binarized images without resolution reduction of the original video frames sampled overtime (details in Figure S2 as supplementary information). Error bars on graphs are based on the statistics of 4 different regions along the dendrite from electrode to center. The extrapolated three dimensional volume was calculated by assuming projected branches to have cylindrical structures, with the cylindrical diameter being equal to the branch width and the cylindrical height being equal to the branch length. The fractal dimension was obtained by box-counting algorithm by counting number of boxes and it quantifies the rate at which an object's geometrical features develop at increasing resolution. For asymmetry determination, the image analysis process was individually performed on each dendrite at both ground and signal electrodes to compare their corresponding density, branches and fractal dimension. Inter-dendrite spacing was determined over time considering their extreme points from both sides, and completion time was defined as the spacing becomes zero. The overall discussion on the two-dimensional image analysis for the three-dimensional dendritic growths acknowledges that the method does not extrapolate three-dimensional physical properties, but serves as a systematic technique to reliably quantify two-dimensional features of three-dimensional objects.

**Cyclic Voltammetry** was performed on a three electrodes setup, with the 25 µm-diameter gold wires as working and counter electrodes, and a macro Ag/AgCl reference electrode. Measurements were performed distinctively for the solutions containing: 1 mM $NaPSS_{(aq)}$, 10 mM $BQ_{(aq)}$ and 1 mM $NaPSS_{(aq)}$, 10 mM $EDOT_{(aq)}$ and 1 mM $NaPSS_{(aq)}$. For each experiment, a new pair of gold wires was used.

**Electrochemical impedance spectroscopy** (EIS) was performed on two gold wire systems with a Solartron Analytical (Ametek) impedance analyzer from 1 MHz to 1 Hz. Different gold electrode systems where characterized with and without dendrite functionalizations.

Systems without dendrite were characterized by voltage-ramped impedance spectroscopy ($V_{DC}$ from 0 to 1.3 V at 1 mV/s, $V_a$ =20 $mV_{rms}$) in the same electrochemical conditions as for the dendritic growths: 10 mM $BQ_{(aq)}$,10 mM $EDOT_{(aq)}$ and 1 mM $NaPSS_{(aq)}$. During the characterization, we observed by microscopy in operando the anodic growth by the appearance of a black coating on the whole surface of the polarized wire. The ePEDOT-coated gold electrode was used to study the electrochemical influence of the ePEDOT material on the impedance of pair of gold wire: the coated wires where re-used as anode, cathode and/or both on subsequent EIS characterization (details in Figure S3 as supplementary information). The thickness of the coatings where uncontrolled and not evaluated.

Systems with dendrite functionalizations were studied as two-electrode systems by impedance spectroscopy with a constant bias ($V_{DC}$= 100mV, $V_a$ =20 $mV_{rms}$) in an electrochemically inactive aqueous electrolyte containing only 1 mM $NaPSS_{(aq)}$, to not promote further electropolymerization of ePEDOT on the dendritic objects *in operando* characterization. In all three cases, dendrites where grown before completion and *a priori*, in the same electrochemical conditions as previously described. Three different morphologies where characterized: one pair of dendrites that are wire-like, one where dendrites are is branchy and vicinal and one where dendrites are branchy and distant (details in Figure S4 as supplementary information).

**Circuit Impedance Modeling** was performed using an open-source EIS Spectrum Analyzer software.[**Bondarenko2005**] The fitting was realized on the raw data spectra without digital-filter preprocessing. The RC parameter fitting was manually adjusted at the visual appreciation of the simultaneous comparison of the Nyquist plots, Bode's modulus and Bode's phase plots.

**Electrical Circuit Simulation** was performed using the Quite Universal Circuit Simulator (QUCS) open-source software,[**QUCS**] based on a (R|C)+R+(R|C) model, with voltage-independent parameters (ideal ohmic resistors and capacitors).

# Acknowledgements

The authors wish to thank the European Commission: H2020-EU.1.1.ERC project IONOS (# GA 773228).

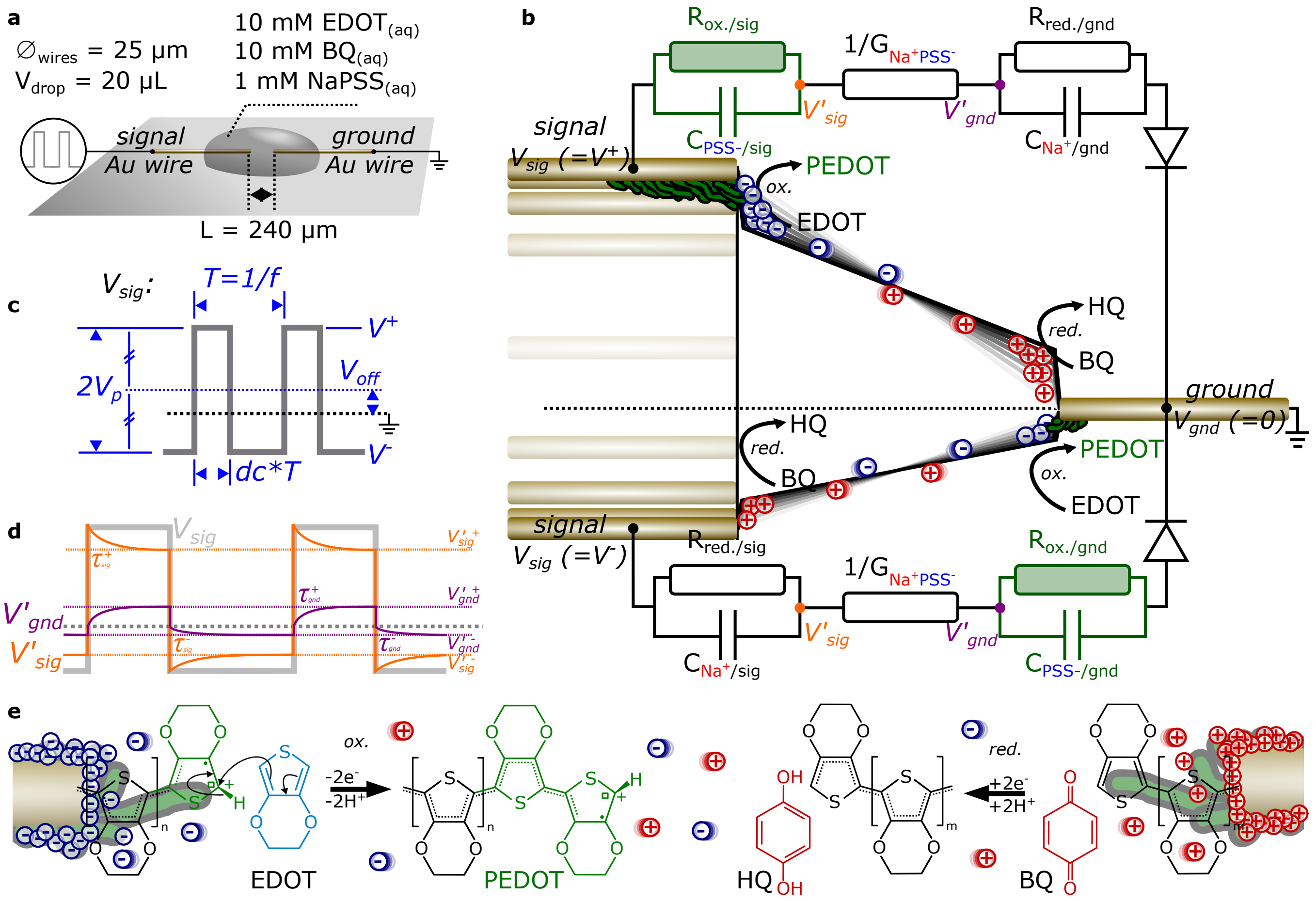

**Figure 1 | Working principle for the growth of conducting polymer dendrites under pulse voltage polarization. a,** Experimental setup to grow dendritic PEDOT on free standing Au wires, polarized under square voltage. **b,** Equivalent circuit of the electrochemical system **c,** Definition of the pulse voltage parameters investigated in this study. **d,** Voltage diagram displaying the ion dynamics, inducing a time dependent voltage drop across the electrolyte, that results in different growth modes for the electropolymerized PEDOT dendrites. **e,** Faradaic (redox) and Non-Faradaic (capacitive) processes occuring at both anodic and cathodic dendritic electrodes upon $V_{sig}$ polarization.

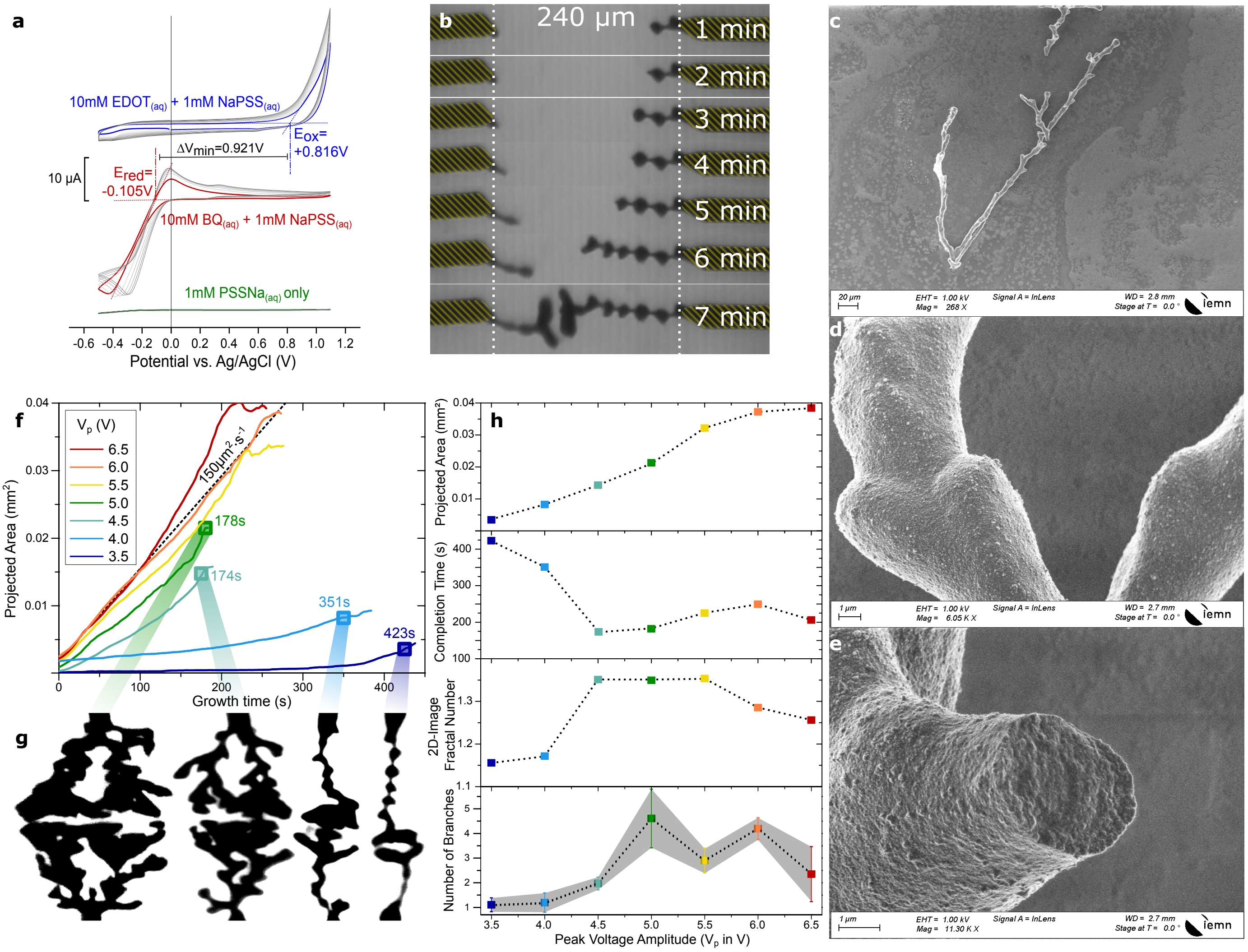

**Figure 2 | ePEDOT Dendrites and their Morphological Control by the Voltage Amplitude ($V_p$). a,** Cyclic Voltammetry for the two Au-wire 1 mM $NaPSS_{(aq)}$ electrochemical system and its behaviour without electrochemically active specie, with 10 mM $EDOT_{(aq)}$ and with 10 mM $BQ_{(aq)}$. **b,**Microscope pictures of one ePEDOT dendritic formation at different time lapses of electropolymerization ($V_p$= 3.5V, $V_{off}$ = 0V, f = 80Hz, dc = 50%. The yellow stripped regions represent the position of the two gold wires (identified from their position at t=0s), in a fresh 10mM $EDOT_{(aq)}$ + 10mM $BQ_{(aq)}$ + 1mM $NaPSS_{(aq)}$ electrolyte. c-e, Scanning Electron Microscope images of an ePEDOT dendritic structure after extraction from its electrolyte. f, Projective dendrite area over time for different dendrites grown at different $V_p$ ($V_{off}$ = 0V, f = 80Hz, dc = 50%). **g,** Contrasted microscope images of four different dendrites, grown at different $V_p$ ($V_{off}$ = 0V, f = 80Hz, dc = 50%), taken at their completion times (highlighted with square marks in **f**). **h,** Relationship between $V_p$ and projective area, completion time, their image's fractal number and the apparent number on branches for dendrites grown up to completion time ($V_{off}$ = 0V, f = 80Hz, dc = 50%).

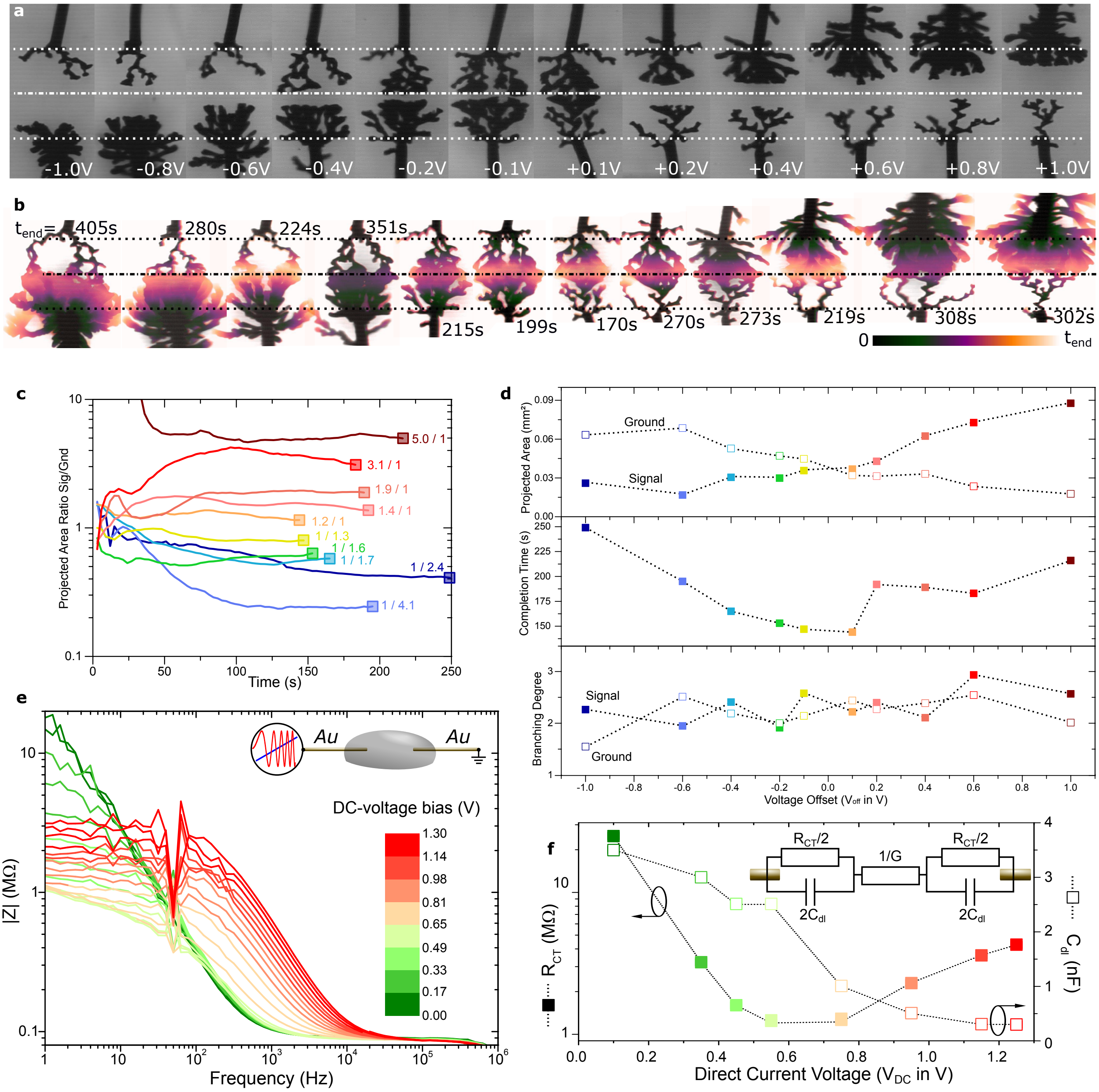

**Figure 3 | Asymmetry Control by the Voltage Offset ($V_{off}$). a,** Microscopes pictures of ePEDOT dendritic formation after t=150 s of voltage polarization under different $V_{off}$ ($V_p$ = 5V, f = 80Hz, dc = 50%). **b,** Color-encoded temporal images of the same dendrites grown from 0 to $t_{end}$ (color scale as inset as linear scale with time). **c,** Ratio between projected area of signal's dendrite on ground's over time, for dendrites grown at different $V_{off}$. **d,** Projected area, completion time and branching degree over $V_{off}$ for dendrites grown up until contact. **e,** Voltage-ramped impedance spectroscopy between $V_{DC}$= 0 and 1.3V of the two gold wire setup immersed in the same electro-active electrolyte (10mM $BQ_{(aq)}$, 10mM $EDOT_{(aq)}$, 1mM $NaPSS_{(aq)}$) that displays a bias-dependent high-pass filter property (the effect of an ePEDOT coating on either or both wires is provided as a supplementary material). **f,** Extracted $R_{CT}$ and $C_{dl}$ parameters from the fitting of the spectrum on the idealized equivalent circuit depicted as an inset (G = 8.5±0.2µS).

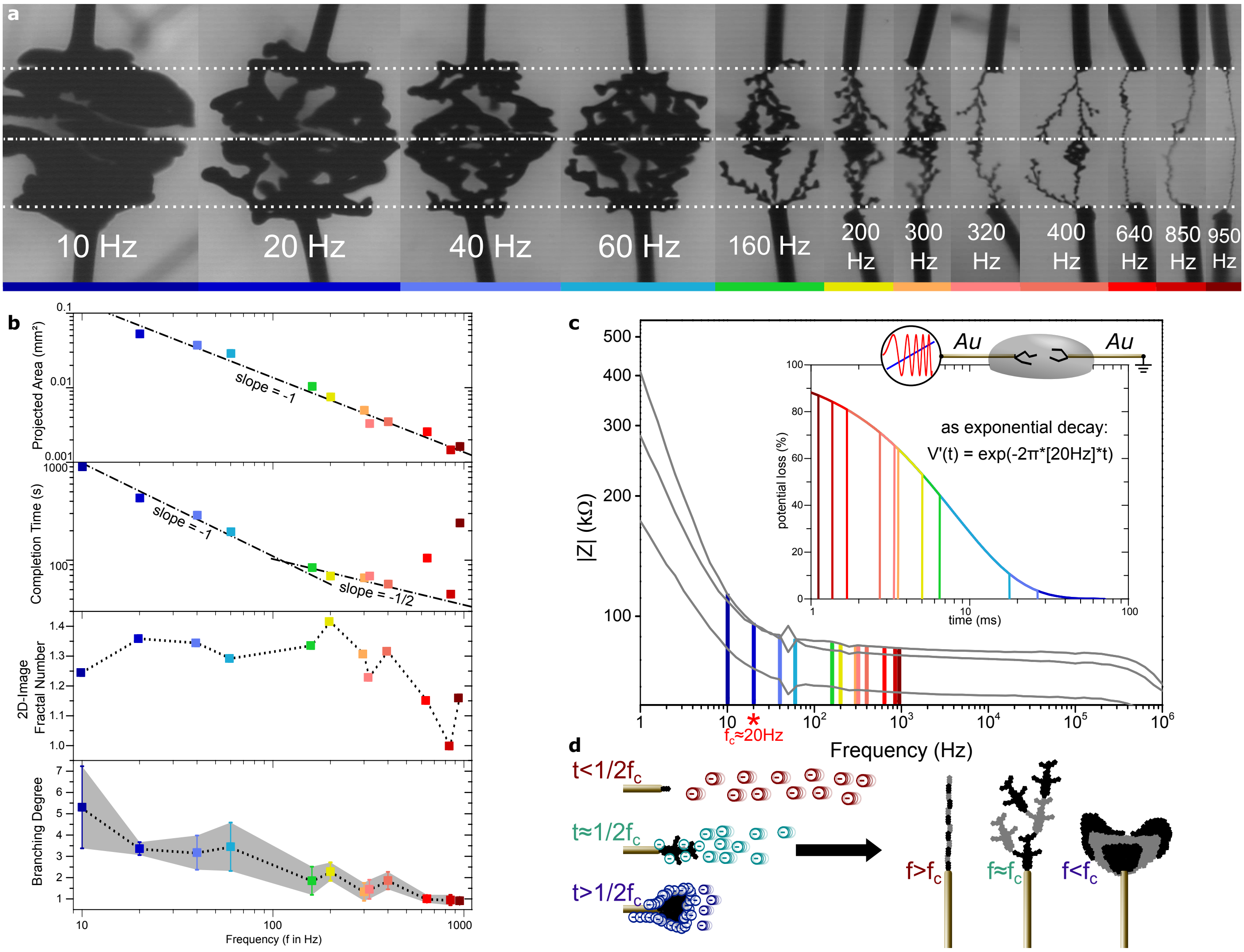

**Figure 4 | Fractal Control by the Frequency (f). a,** Microscope pictures of ePEDOT dendritic formations (signal wire: top, ground wire: bottom) at their completion time grown under various frequencies from 10Hz to 950Hz ($V_p$ = 5V, $V_{off}$ = 0V, dc = 50%). **b,** Projected area, completion time, fractal number of their image and branching degree over f for dendrites grown up until contact. **c,** Impedance spectroscopy of the two-wire setup after three different dendritic formations (see details on the three evaluated dendrites as supplementary information), performed in 1 mM $NaPSS_{(aq)}$ (without $EDOT_{(aq)}$ nor $BQ_{(aq)}$), that displays in the three cases a change in the frequency-dependent regime of the modulus at $f_c$ around 20Hz. Inset: Decay over time of the potential across the wire assuming an ideal resistor-capacitor discharge, and an exponential decay rate $\tau = 1/[2\pi*20Hz]$. **d,** Schematics for the growth of three different dendrites with morphological dependencies associated to anionic screening during polarization responsible for the potential drop at the anode, for which the hindrance time dependency is associated to the frequency and the $f_c$ of the electrochemical system.

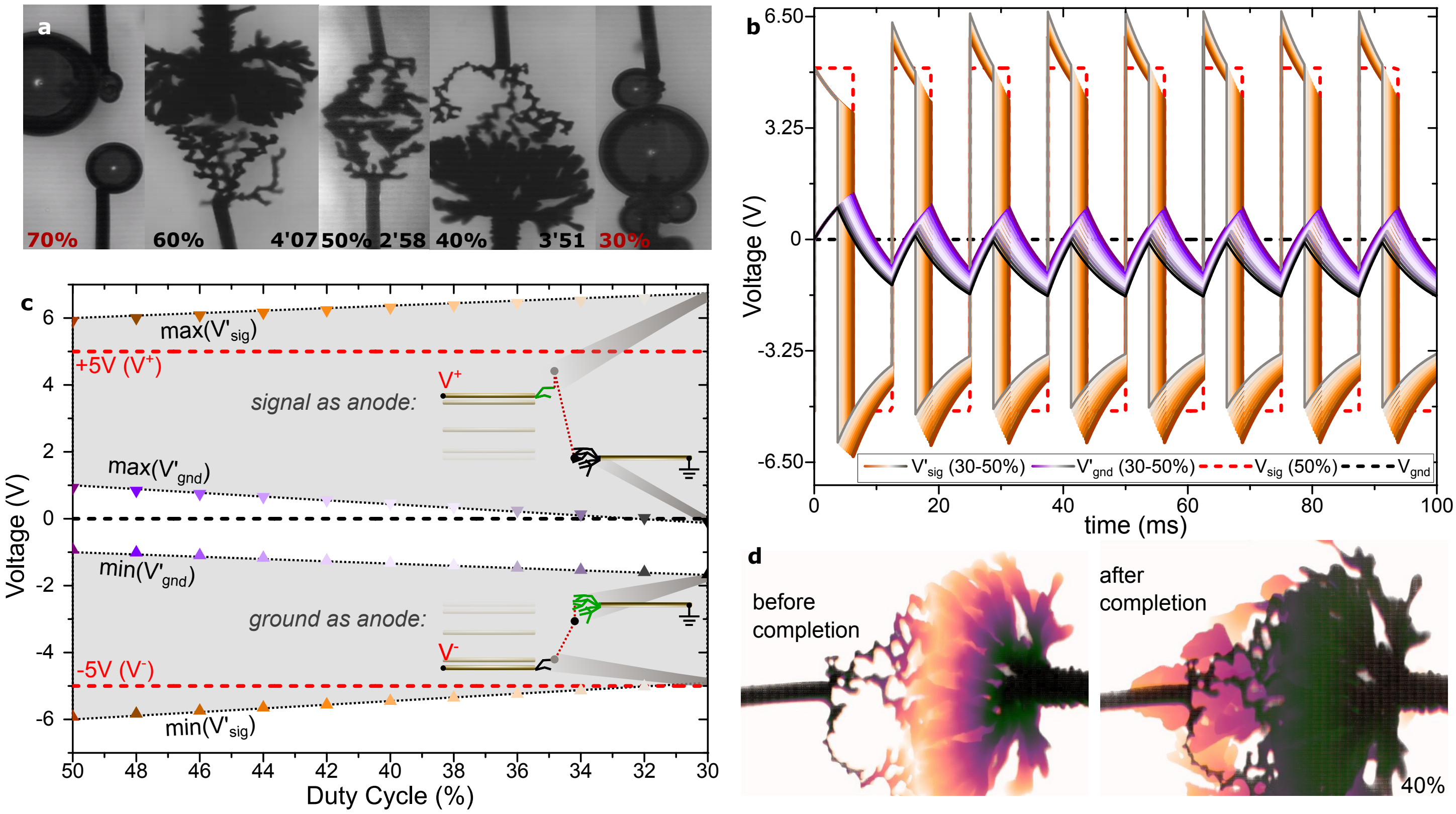

**Figure 5 | Dendritic Engulfing and Asymmetry with the Duty-Cycle (dc). a,** Microscopes pictures of ePEDOT dendritic formations (signal wire: top, ground wire: bottom) at completion time after growth under different duty cycles ($V_p$ = 5V, $V_{off}$ = 0V, f= 80Hz) the two extremes cases dc = 30% and 70% where taken at t = 1min). **b,** Simulation of the $V'_{sig}(t)$ and $V'_{gnd}(t)$ dynamics under a pulse waveform ($V_p$ = 5V, $V_{off}$ = 0V, f = 80Hz) with various duty cycles between 50% and 30% (assuming the equivalent circuit of Fig.4f inset with rigid parameters as $R_{CT}$ = 2MΩ, 1/G = 80kΩ and $C_{dl}$ = 10µF). **c,** Dependency of $V'_{sig}$ and $V'_{gnd}$ voltages at the change of the signal polarity (spike) upon variations of the waveform's duty cycle between 50% and 30% from simulations displayed in Fig. 6c (the two diagrams show the different voltage drops between the dendrite at the both changes of the waveform's polarity for dc = 30%). **d,** Color-encoded images of a dendritic growth at dc = 40% showing two specific dynamics before and after completion.

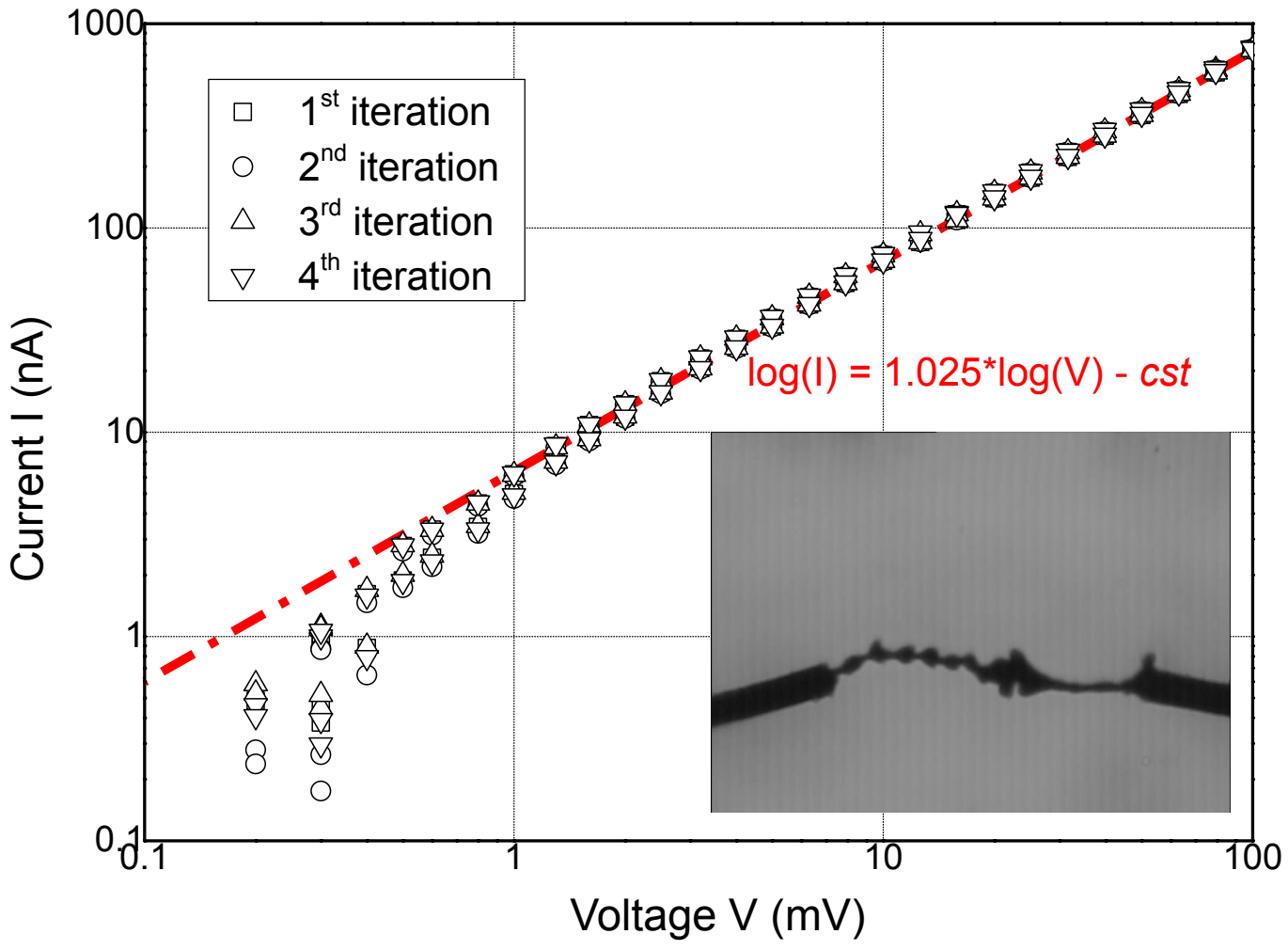

**Figure S1| Electrical Characterization of an ePEDOT dendritic formation a,** Four repeted iterations of output current characterization of the dendrite (displayed as an inset, grown with Vpp = 3.5V, Voff = 0V, f = 80Hs, dc = 50%) upon DC voltage polarization at the wires, showing a Ohmic behaviour for the microstructure (voltage exponent close to one). The *cst* value of 5.1 (I in ampere and V in volt) indicates a conductance for the microstructure of 7.4 µS (resistance of 135 kΩ). Considering the length of the dedritic wire of 240 µm and an apparent section between 2 to 10 µm, the conductivity of the material is evaluated to be between 0.21 and 5.64 S/cm (assuming it to be uniform).

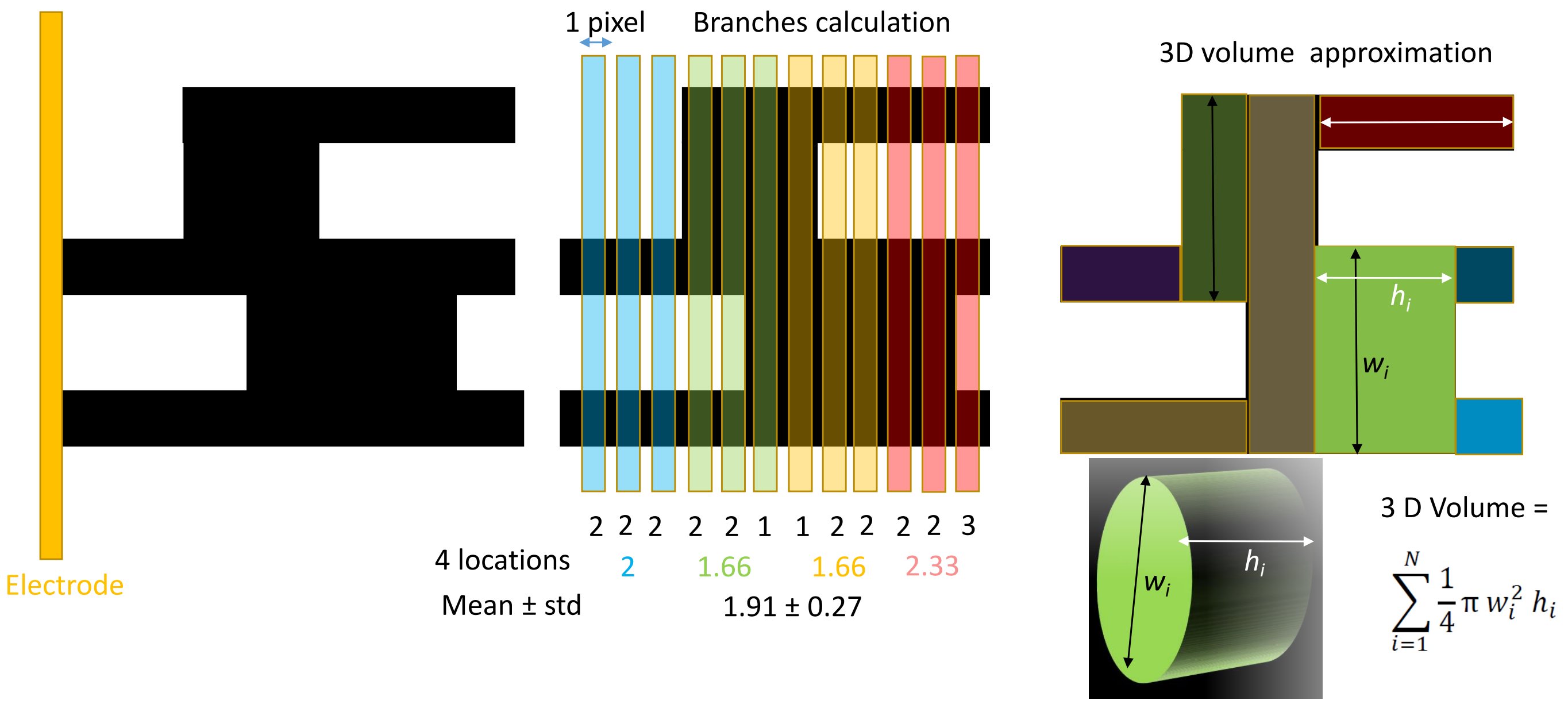

**Figure S2 | Binarized Image Pixel Counting. a,b,** Branching number was determined based on total intersections of dendrite structures with elementary vertical lines as pixels. **c,** Extrapolated three dimensional volume was calculated assuming the dendrite branches to have cylindrical elementary structures, with the cylinder diameter to be equal to the branch width and the cylinder oheight to be equal to the branch length.

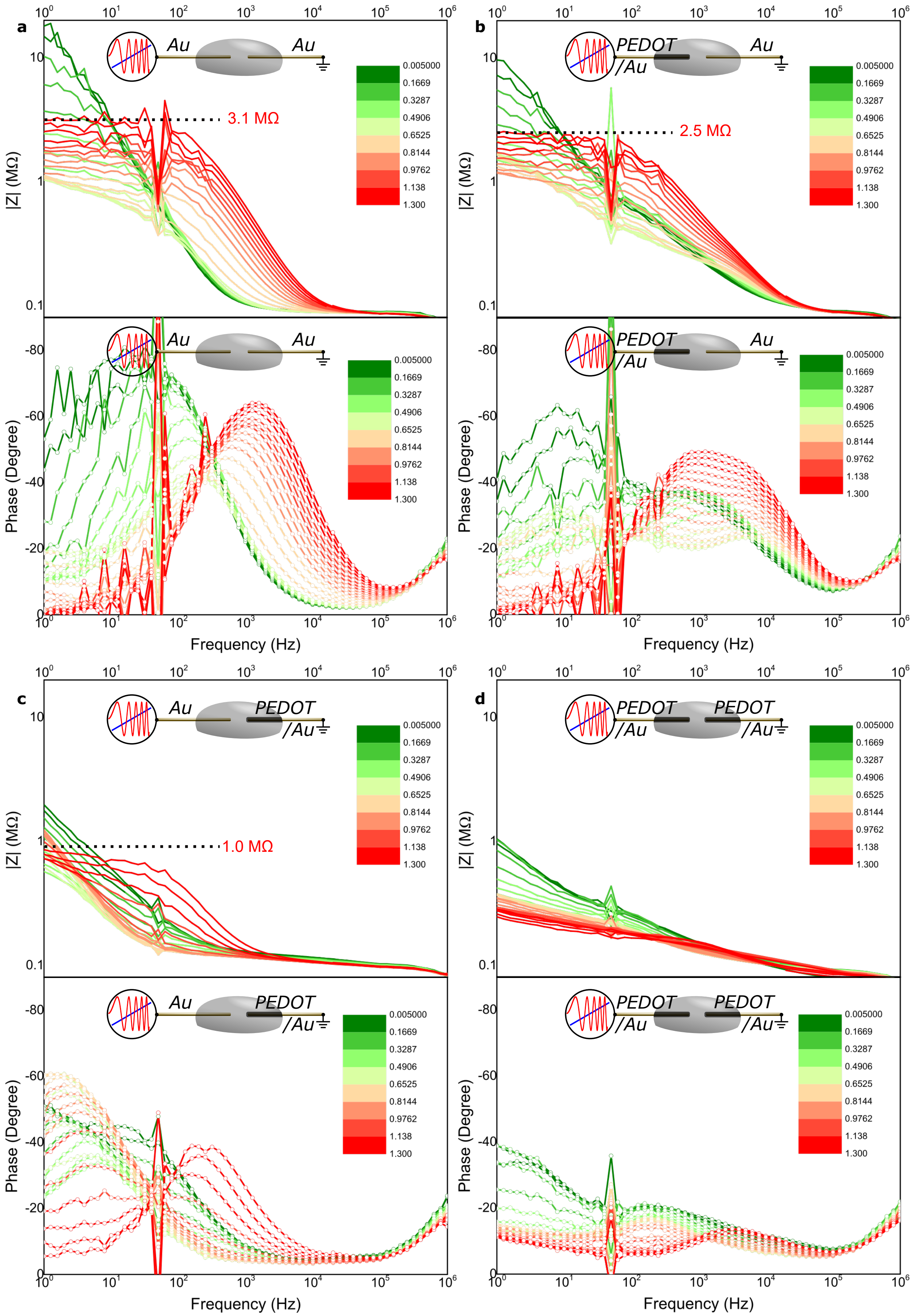

**Figure S3 | Voltage-Ramp Impedance Spectroscopy on Au Wires.** Bode impedance modulus and phase diagrams of naked gold wires (**a**), with a PEDOT coating exclusively on the signal-wire (**b**), or on the grounded wire (**c**), and on both wires (**d**). The spectra were generated in a comparable setup as the studies performed with square waveforms (1 mM $NaPSS_{(aq)}$, 10 mM $BQ_{(aq)}$, 10 mM $EDOT_{(aq)}$, 240 µm) with a DC voltage ramp ($V_{DC}$) from 0 to 1.3 V of 1mV/s (color scale of $V_{DC}$ in volt).

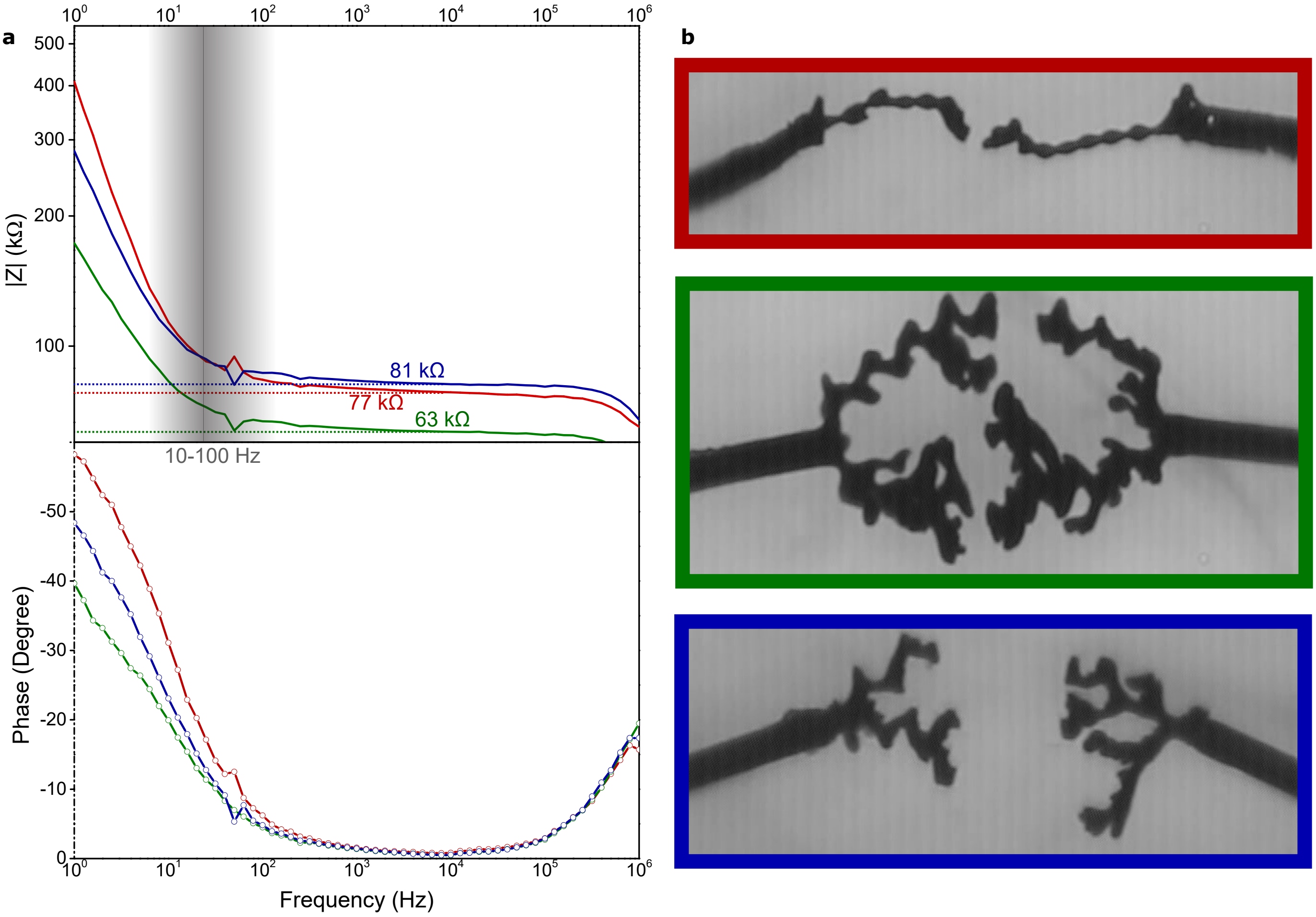

**Figure S4 | VImpedance Spectroscopy on Dendritic Structures.a,** Bode impedance modulus and phase diagrams of dendritic structures grown on gold wires (distance: 240 µm): a narrower-gap & wire-like structures (red), narrower-gap & fractal dendrites (green) and wider-gap & fractal dendrites (blue). b, Microscope picture showing the morphology of the three dendrites characterized by impedance spectroscopy.